\documentclass[a4paper,fleqn,usenatbib]{mnras}
\usepackage[english]{babel}
\usepackage{hyperref}
\usepackage{cleveref}
\usepackage{newtxtext,newtxmath}
\usepackage[T1]{fontenc}
\usepackage{ae,aecompl}
\usepackage[latin9]{inputenc}
\usepackage{graphicx,amsmath,amssymb}
\newcommand{\be} {\begin{equation}}
\newcommand{\ee} {\end{equation}}
\newcommand{\bea}{\begin{eqnarray}}
\newcommand{\eea}{\end{eqnarray}}
\def\aap{A\&A}
\def\apj{ApJ}
\def\aj{AJ}
\def\apjs{ApJ S}
\def\mnras{MNRAS}
\def\cemda{CeMDA}
\def\icarus{Icarus}

\begin{document}
\title[Resonance Capture and Dynamics of 3-Planet Systems]{Resonance Capture and Dynamics of 
3-Planet Systems}
\author[C. Charalambous et al.]{
C. Charalambous,$^{1}$\thanks{E-mail: charalambous@oac.unc.edu.ar}
J.G. Mart\'{\i},$^{2}$
C. Beaug\'e$^{1}$
and X.S. Ramos$^{1}$
\\
$^{1}$Instituto de Astronom\'{\i}a Te\'orica y Experimental, Observatorio Astron\'omico, \\
Universidad Nacional de C\'ordoba, Laprida 854, C\'ordoba X5000GBR, Argentina\\
$^{2}$Instituto de Astrof\'isica de La Plata, Facultad de Ciencias Astron\'omicas y Geof\'isicas, \\
Universidad Nacional de La Plata, La Plata, Argentina
}
\date{}
\pubyear{2017}
\label{firstpage}
\pagerange{\pageref{firstpage}--\pageref{lastpage}}
\maketitle

\begin{abstract}
We present a series of dynamical maps for fictitious 3-planets systems in initially circular 
coplanar orbits. These maps have unveiled a rich resonant structure involving two or three 
planets, as well as indicating possible migration routes from secular to double resonances or 
pure 3-planet commensurabilities. These structures are then compared to the present-day 
orbital architecture of observed resonant chains. 
In a second part of the paper we describe N-body simulations of type-I migration. Depending on 
the orbital decay timescale, we show that 3-planet systems may be trapped in different 
combinations of independent commensurabilities: (i) double resonances, (ii) intersection between 
a 2-planet and a first-order 3-planet resonance, and (iii) simultaneous libration in two 
first-order 3-planet resonances. These latter outcomes are found for slow migrations, while 
double resonances are almost always the final outcome in high-density disks. 
Finally, we discuss an application to the TRAPPIST-1 system. We find that, for low migration 
rates and planetary masses of the order of the estimated values, most 3-planet sub-systems are 
able to reach the observed double resonances after following evolutionary routes defined by pure 3-planet resonances. The final orbital configuration shows resonance offsets comparable 
with present-day values without the need of tidal dissipation. For the 8/5 resonance proposed 
to dominate the dynamics of the two inner planets, we find little evidence of its dynamical 
significance; instead, we propose that this relation between mean motions could be a consequence 
of the interaction between a pure 3-planet resonance and a 2-planet commensurability 
between planets \texttt{c} and \texttt{d}. 
\end{abstract}


\section{Introduction}
\label{sec:intro}

TRAPPIST-1 \citep{Gillon.etal.2016, Gillon.etal.2017, Luger.etal.2017} is a unique exoplanetary 
system of seven planets in a complex resonant chain comprised of five interlocked zero-order 
(i.e. Laplace) 3-body mean-motion resonances. Although the multi-resonant state is not yet 
confirmed and most initial conditions consistent with the observations lead to dynamical 
instabilities in short timescales (\citealp{Gillon.etal.2017}), N-body simulations by 
\cite{Tamayo.etal.2017} indicated that similar stable configurations may be reached by smooth 
planetary migration. 

In recent years several transit systems have been discovered in multi-planet resonances: 
Kepler-60 \citep{Steffen.etal.2013, Gozdziewski.etal.2016}, Kepler-80 
\citep{MacDonald.etal.2016} and most noticeably Kepler-223 \citep{Mills.etal.2016} where 
precise TTVs spanning over 4 years of observations have shown the actual libration of the 
Laplace angles. Independently of the known number of planets, the fundamental building blocks of 
all these resonance chains consist of 3-body Laplace resonances. Thus, independent of the planet 
multiplicity, many dynamical properties of multi-resonant systems may be tackled by studying 
3-planet commensurabilities. 

A particularly interesting case is Kepler-444 \citep{Campante.etal.2015}, with five planets 
orbiting the host star within $0.8\,au$. This is a noteworthy system for two reasons. First, the 
central star has a stellar companion at $\sim 60\, au$, making the binary sufficiently tight to 
have influenced the dynamical evolution and, possibly, the formation process itself. Second, 
the age of host star is estimated at 11.2 Gyrs, one of the oldest planetary systems known to 
date, and thus, a good candidate to analyze how tidal effects may have altered its primordial 
orbital architecture.

3-planet resonances may also play a relevant role in defining the orbital architecture of 
our own solar system. Apart from the classical Laplace resonance between the inner three 
Galilean satellites \citep[e.g.][]{Yoder.1979}, resonant chains have also been proposed as 
acting between the outer planets \citep{Murray.Holman.1999, Guzzo.2005, Guzzo.2006} possibly 
leading to (extremely weak) chaotic motion in the outer solar system. 
 
In this work we will present a series of dynamical maps of the orbital-period ratio 
representative plane of initial conditions for 3-planet systems. These will help unveil the 
complex richness of resonant structures as well as the relative strengths and common origin 
between 2-planet and several different types of 3-planet resonances. In a second part, we 
study the migration and resonance capture of the TRAPPIST-1 system and analyze outcomes for 
fictitious systems as a function of the initial conditions, planetary masses and migration 
rates. Finally, a discussion is presented in how different types of resonant configurations may 
be reached in each case.

\section{The Dynamical System}

\subsection{Variables}

Our dynamical system consists of three planetary masses $m_i$ ($i = 1,2,3$) in coplanar orbits 
around a central star of mass $m_0$, with $m_0 >> m_i$. We will denote with $a_i$ the 
semimajor axes, $e_i$ the eccentricities, $\lambda_i$ the mean longitudes and $\varpi_i$ the 
longitudes of pericenter of each planet. All orbital elements are defined in a Jacobi reference 
frame.

Since our analytical model will be based on a Hamiltonian formalism, it is useful to first 
introduce the modified Delaunay canonical variables which, in the planar problem, are given by:
\be
\begin{split}
L_i = m_i' \sqrt{\mu_i a_i} \hspace*{1.6cm} &; \hspace*{0.45cm} \lambda_i  \\
S_i = L_i \left( 1-\sqrt{1-e_i^2} \right) \hspace*{0.5cm} &; \hspace*{0.2cm} 
	-\varpi_i  \\
\end{split}
\label{eq.1}
\ee
where the mass factors acquire the form:
\be
\mu_i = {\cal G} \sum_{j=0}^i m_j \hspace*{0.6cm} ; \hspace*{0.6cm} 
m'_i = m_i \frac{\sum_{j=0}^{i-1} m_j}{\sum_{j=0}^{i}  m_j} ,
\label{eq.2}
\ee
the latter being the reduced mass of the $i$-th planet. The gravitational constant is denoted 
by ${\cal G}$. 

The Hamiltonian $F$ for the system can then be written as the sum of two terms $F = F_0 + F_1$; 
the first leads to the Keplerian motion of the planets around the central star, while $F_1$ 
groups all perturbations arising from mutual gravitational interactions between the planets.
Written in the Delaunay variables \eqref{eq.1}, the integrable Hamiltonian $F_{0}$ acquires de 
form:
\be
F_{0} = -\sum_{i=1}^N \frac{\mu_i^2 m_i'^3}{2L_i^2} = -  \frac{\mu_1^2 m_1'^3}{2L_1^2} - 
\frac{\mu_2^2 m_2'^3}{2L_2^2} - \frac{\mu_3^2 m_3'^3}{2L_3^2}, 
\label{eq.3}
\ee
while the perturbation term can be generically expressed as:
\be
F_{1} \equiv -{\cal R} = -{\cal R}_{12} - {\cal R}_{23} - {\cal R}_{13} ,
\label{eq.4}
\ee
where ${\cal R}_{ij}$ denotes the disturbing function that arises from the interaction between 
$m_i$ and $m_j$. Retaining only terms corresponding to the lowest order of the masses, the 
gravitational perturbations have the same functional form as the one deduced for the restricted 
three-body problem \citep[e.g.][]{Libert.Henrard.2007a}. Thus, the planetary disturbing function 
may be expressed in terms of the position vectors ${\vec r}_i$ as:
\be
{\cal R}_{ij} = {\cal G} m_i m_j \biggl( \frac{1}{|{\vec r}_i - {\vec r}_j|} - 
\frac{{\vec r}_i \cdot {\vec r}_j}{|{\vec r}_j|^3} \biggr),
\label{eq.5}
\ee
where ${\vec r}_i$ are in the Jacobi reference frame. 

\subsection{Transformation Between Mean and Osculating Elements}
\label{transf}

All resonant conditions are defined in mean variables (i.e. averaged over short period terms)
but our dynamical maps will be calculated in a representative plane of osculating elements. 
While the difference between both sets may not be significant for low planetary masses and/or 
for systems far from the Hill stability limit \citep[e.g.][]{FerrazMello.etal.2005ApJ, 
Deck.etal.2013, Ramos.etal.2015}, it will prove important to correctly identify the resonances 
appearing in the dynamical maps and to estimate their relative strength. 

Although the details of the canonical transformation for 2-planet systems are well documented
\citep[e.g.][]{Tisserand.1889, Deck.etal.2013, Ramos.etal.2015}, the extension to 3-planet 
systems are not easily available and will be given here. The steps for the construction of the 
generating function are analogous, the only significant difference is the existence of three 
independent terms in the disturbing function \eqref{eq.5}. We follow, thus, the procedure 
employed by \cite{Ramos.etal.2015} extended to the case of three planets. 

Let us denote by $(L_i,S_i,\lambda_i,-\varpi_i)$ the osculating set of variables, while the 
mean canonical elements will be expressed by $(L^*_i,S^*_i,\lambda^*_i,-\varpi^*_i)$. We then 
search for a Lie-type generating function  $B: (L_i,S_i,\lambda_i,-\varpi_i) 
\rightarrow (L^*_i,S^*_i,\lambda^*_i,-\varpi^*_i)$ such that the transformed Hamiltonian is 
independent of the new mean longitudes, i.e. $F^\star = F^\star(S^*_i,-\varpi^*_i;L^*_i)$. Up to 
lowest order in the masses, the relation between both sets of variables will be explicitly given 
by:
\be
\begin{split}
L_i &= L^*_i + \frac{\partial B}{\partial \lambda_i} \hspace*{0.5cm} ; \hspace*{0.5cm}
\;\; \lambda_i = \lambda^*_i - \frac{\partial B}{\partial L_i}  \\
S_i &= S^*_i - \frac{\partial B}{\partial \varpi_i} \hspace*{0.5cm} ; \hspace*{0.5cm}
\varpi_i = \varpi^*_i + \frac{\partial B}{\partial S_i} ,
\end{split}
\label{eq.6}
\ee
with $(i=1,2,3)$. The first-order generating function $B$ is the solution of the partial 
differential equation
\be
- \boldsymbol{n} \cdot \frac{\partial B}{\partial \boldsymbol{\lambda}} = \{ F_1 \}
= -\{ {\cal R}_{12} \} - \{ {\cal R}_{23} \} - \{ {\cal R}_{13} \}
\label{eq.7}
\ee
where $\boldsymbol{n} = (n_1,n_2,n_3)$ is the mean-motion vector, $\boldsymbol{\lambda} = 
(\lambda_1,\lambda_2,\lambda_3)$ is the mean longitudes vector, and $\{ {\cal R}_{ij} \}$ 
denotes the purely periodic part of ${\cal R}_{ij}$ \citep[e.g.][]{Hori.1961, FerrazMello.2007}. 
Since we have chosen to express $B$ in osculating variables, the transformation equations 
\eqref{eq.6} will have to be solved iteratively; however, the precision gained by this approach 
makes the extra work worthwhile.

In the case of circular orbits, and neglecting the indirect terms, the Laplace expansion of the 
disturbing function acquires the form (e.g. \citealp{Brouwer.Clemence.1961}) 
\be
{\cal R}_{ij} = \frac{{\cal G} m_i m_j}{2 a_j} \sum_{k=-\infty}^{\infty} b^{(k)}_{1/2} 
                (\alpha_{ij}) \cos{k(\lambda_i - 
\lambda_j)} 
\label{eq.8}
\ee
where $\alpha_{ij} = a_i/a_j$ are the semimajor-axes ratios. Adding up the three different 
gravitational functions, eliminating the secular (i.e. non-periodic) terms, and introducing the 
resulting expression into \eqref{eq.7}, we can explicitly calculate the generating function, 
yielding
\be
\begin{split}
B &= \frac{{\cal G} m_1 m_2}{a_2(n_1-n_2)} \sum_{k=1}^{\infty} \frac{1}{k} 
        b^{(k)}_{1/2}(\alpha_{12}) \; \sin{k(\lambda_1 - \lambda_2)} \;\; + \\
  &+ \frac{{\cal G} m_2 m_3}{a_3(n_2-n_3)} \sum_{k=1}^{\infty} \frac{1}{k} 
        b^{(k)}_{1/2}(\alpha_{23}) \; \sin{k(\lambda_2 - \lambda_3)} \;\; + \\
  &+ \frac{{\cal G} m_1 m_3}{a_3(n_1-n_3)} \sum_{k=1}^{\infty} \frac{1}{k} 
        b^{(k)}_{1/2}(\alpha_{13}) \; \sin{k(\lambda_1 - \lambda_3)}.
\end{split}
\label{eq.9}
\ee

Since we have adopted circular orbits, $B$ does not depend on either $S$ or $\varpi$, leading 
to identical values in both sets of variables. Moreover, choosing initial conditions with 
$\lambda_i=0$ also leads to $\lambda^*_i = 0$, while the change in the Delaunay action $L_i$ 
may be written in terms of the original form of the disturbing functions as:
\be
\begin{split}
L_1 = L^*_1 & + \frac{\{ {\cal R}_{12} \} }{(n_1-n_2)} 
              + \frac{\{ {\cal R}_{13} \} }{(n_1-n_3)}   \\
L_2 = L^*_2 & - \frac{\{ {\cal R}_{12} \} }{(n_1-n_2)} 
              + \frac{\{ {\cal R}_{23} \} }{(n_2-n_3)}  \\
L_3 = L^*_3 & - \frac{\{ {\cal R}_{13} \} }{(n_1-n_3)}  
			  - \frac{\{ {\cal R}_{23} \} }{(n_2-n_3)} .
\end{split}
\label{eq.10}
\ee
Since $\lambda_i=0$, the amplitude of the periodic functions can be written explicitly as
\be
\begin{split}
\{ {\cal R}_{ij} \} &= \frac{{\cal G} m_i m_j}{a_j} \biggl( \frac{a_j}{\Delta_{ij}}
     - \frac{1}{2} b^{(0)}_{1/2} (\alpha_{ij}) \biggr) \\
     &= \frac{{\cal G} m_i m_j}{a_j} \biggl( \frac{1}{(1-\alpha_{ij})} 
     - \frac{1}{2} b^{(0)}_{1/2} (\alpha_{ij}) \biggr) ,
\end{split}     
\label{eq.11}
\ee
where $\Delta_{ij} = |{\vec r}_i - {\vec r}_j|$. Introducing this expression into \eqref{eq.10}, 
we can obtain closed analytical formula for the transformation between mean and osculating 
Delaunay momenta and, consequently, between the semimajor axes. 

\begin{figure}
\centering
\includegraphics*[width=\columnwidth]{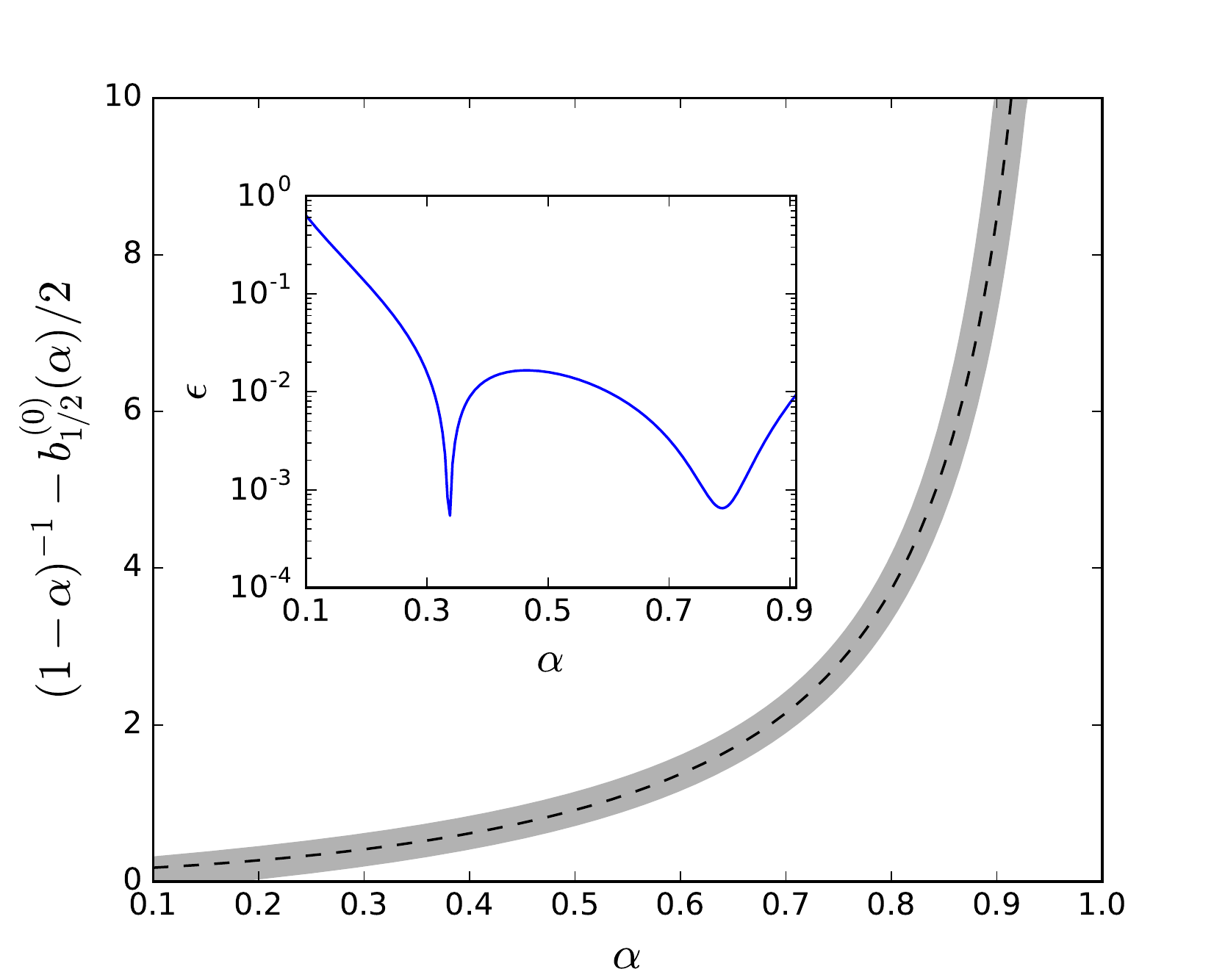}
\caption{Broad gray curve in the main plot shows the function $(1-\alpha)^{-1} - (1/2)
b^{(0)}_{1/2}(\alpha)$ as function of the semimajor axis ratio, while the thin dashed curve 
corresponds to the approximation described in equation \eqref{eq.12}. The inlaid graph plots
the relative error $\epsilon$ of the minimum-squares fit.}
\label{fig.b0_12}
\end{figure}

However, an additional simplification may be performed. Inspired by the surprising linear 
correlation found by \citep{Wisdom.1980} between the amplitude of the main resonant term and the 
degree $p$ of a given first-order resonance, we searched for a similar trend for the 
short-period terms. Although we failed to find a similar expression for the Laplace coefficient 
itself, we did find a suitable approximation for the full amplitude of the short-period 
perturbation \eqref{eq.11}. Explicitly we found that
\be 
\frac{1}{1-\alpha_{ij}} - \frac{1}{2}\, b^0_{1/2}(\alpha_{ij}) \simeq 1.43\, p_{ij} + 0.13 \,,
\label{eq.12}
\ee
where the numerical coefficients were determined using a least-squares linear fit in the
parameter $p_{ij}$, defined as
\be 
p_{ij} = \frac{\alpha_{ij}^{3/2}}{1-\alpha_{ij}^{3/2}}.
\label{eq.13}
\ee
When initial conditions place the semimajor axis ratio in a first-order resonance, then 
$p_{ij}$ is an integer and equal to the degree of that commensurability, i.e. 
$n_i/n_j = (p_{ij}+1)/p_{ij}$. In any other configuration $p_{ij}$ takes non-integer values. 
Figure \ref{fig.b0_12} compares the predictions of \eqref{eq.12} (dashed curve) with the exact 
values (broad gray line), while the inlaid plot highlights the relative error $\epsilon$ between 
both. The linear fit guarantees a maximum relative error $\epsilon \sim 10^{-2}$ for all orbital 
separations in the interval $\alpha \in [0.3,0.9]$.

\section{Resonant Structure}
\label{sect:res_structure}

\subsection{Dynamical Maps}

We begin with a numerical study of the resonant structure of the 3-planet problem. This will be 
accomplished by means of a series of dynamical maps in the $(n_1/n_2,n_2/n_3)$ plane, with 
initial conditions corresponding to circular planar orbits with all angles equal to zero. 
This choice corresponds to a collinear configuration where the mutual distance between the 
planets is minimum. 

In our numerical simulations, we integrated the equations of motions of the four bodies of the 
system (central mass plus three planets) in a Jacobi reference frame, using a Bulrisch-St\"oer 
algorithm with a precision specified by a maximum permitted relative error per time-step of 
$10^{-13}$. We set the central mass to $m_0 = 1 m_\odot$ and chose the initial $a_3$ to be 
always equal to $1\,au$. Each initial configuration with different initial semimajor axis 
ratios was integrated for a total time span of $T = 10^4$ years (which in this case represent 
the total orbits of the outer body). During the integrations we kept track of the variation of 
each planet's semimajor axis, being able to calculate at the end each planet's maximum 
variation during the whole timespan, $\Delta a_i = ({a_i}_{\rm max} - {a_i}_{\rm min})$ 
\citep[e.g.][]{Gallardo.etal.2016}. We also calculated for each initial condition the maximum
value of $\Delta a$, which is the maximum of the planetary variations: $max(\Delta a) = 
{\rm max}(\Delta a_1,\Delta a_2,\Delta a_3)$. Although this indicator does not measure 
chaotic motion, it is useful for mapping the resonant structure and analyzing the behavior of 
planetary systems, very similar to the better known maximum eccentricity method 
\citep[e.g.][]{Dvorak.etal.2004, Ramos.etal.2015}. The $max(\Delta {a})$ measure was chosen 
over its $max(\Delta {e})$ counterpart since it better identifies Laplace-type resonances, 
where the eccentricity suffers no appreciable excitation. 

\begin{figure}
\centering
\includegraphics*[width=0.95\columnwidth]{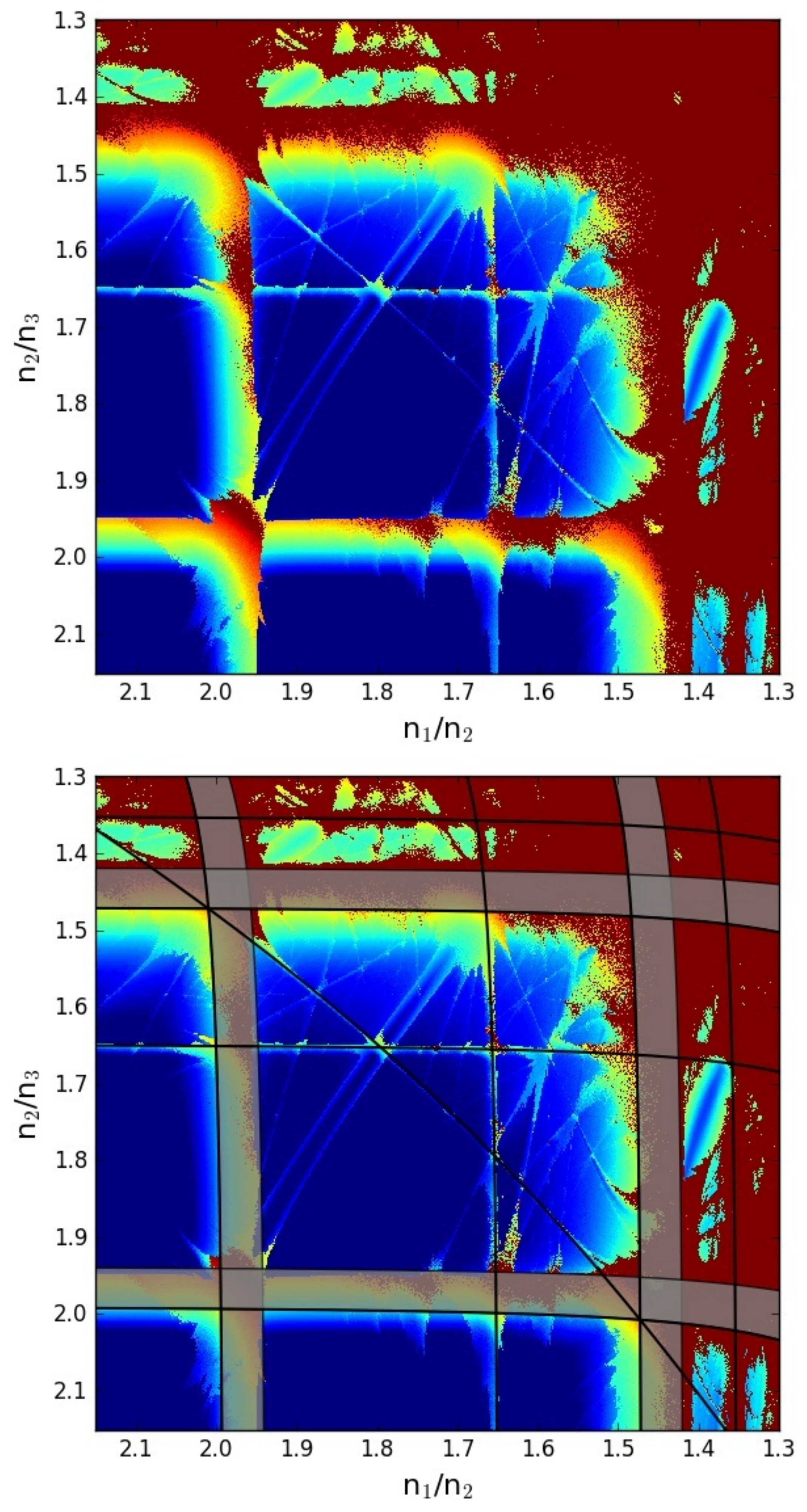}
\caption{{\bf Top:} Dynamical map of $max(\Delta a)$ in the $(n_1/n_2,n_2/n_3)$ plane for a grid 
of $(800 \times 800)$ initial conditions in circular coplanar orbits. Blue tones indicate low 
changes in semimajor axes, while yellow and red tones indicate increasing values. Total 
integration times was $T = 10^4$ orbits of the outer planet. We considered equal-mass planets 
with $m_1=m_2=m_3=150 m_{\oplus}$ orbiting a central star with $m_0 = 1 m_\odot$. {\bf Bottom:} 
Superimposed to the dynamical map, black lines show the nominal location of relevant 2-planet 
mean-motion resonances while for first-order commensurabilities the distance to the inner 
separatrix are marked by gray regions. The diagonal curve corresponds to the 3/1 MMR between 
$m_1$ and $m_3$.}
\label{fig.2b-mmr}
\end{figure}

The top frame of Fig.~\ref{fig.2b-mmr} shows a dynamical map calculated over a grid of $800 
\times 800$ initial conditions where all three masses were taken equal to $m_i = 150 \, 
m_{\oplus}$. The color code indicates the value of $max(\Delta a)$ after $T=10^4$ yrs. Blue 
corresponds to small changes in the semimajor axes (usually indicative of regular motion), while 
red indicates large variations. These may correspond either to dynamically unstable orbits 
(escapes or collisions) or to stable initial conditions close to resonant separatrix, whose 
dynamics led to high eccentricities. The integration time was chosen sufficiently large to map 
the main features of the resonant structure but not so long so as to blur them with chaotic 
diffusion. Thus, at this point we are more interested in mapping the phase space than in 
identifying stable/unstable domains. 

The phase plane shows a rich structure generated by a web of two and three-planet resonances. 
All commensurabilities in this plane are characterized by a condition of type
\be
j_1 n_1 + j_2 n_2 + j_3 n_3 \simeq 0,
\label{eq.14}
\ee
for some set of $(j_1,j_2,j_3) \neq (0,0,0)$. Since resonance relations are defined in 
mean orbital elements while the dynamical maps are constructed from a grid of initial conditions 
in osculating elements, we must use the transformation equations deduced in the previous 
section to relate both sets of variables. 

\subsection{2-Planet Mean-Motion Resonances (2P-MMR)}

Superimposed to the dynamical map, the bottom frame of Fig.~\ref{fig.2b-mmr} shows the main 
features of two-planet mean-motion resonances (hereafter 2P-MMR). Commensurabilities between 
$m_1$ and $m_2$ appear as almost vertical curves, where the curvature is caused by the fact that 
we are plotting osculating elements and not their mean counterparts. The functional form of the 
curves were calculated from the expressions deduced in the previous section and are mainly 
caused by short-period perturbations from the non-resonant third planet (in this case, $m_3$). 
From left to right we observe the 2/1, 5/3, 3/2 and 7/5 MMRs, whose nominal location is 
identified by broad black curves. The observed shift with respect to the exact commensurability 
relations is this time due to the short-period perturbations between both $m_1$ and $m_2$.

While second-order MMRs have negligible libration widths for circular orbits, first-order 
resonances cause a significant change in both eccentricity and semimajor axis for all initial 
conditions between the nominal resonant value and the inner separatrix at $e_i=0$ 
(see \citealp{Ramos.etal.2015}). This region is shaded in gray, where the semi-width of the 
libration region was estimated using the analytical expression by \cite{Deck.etal.2013}. 

The same 2P-MMRs, now between $m_2$ and $m_3$, are depicted as near-horizontal curves. Once 
again the broad black curves correspond to the nominal location while, the regions up to the 
inner separatrix are shown in gray. The structures associated to both two-planet resonances are 
symmetric with respect to the diagonal line defined by $n_1/n_2 = n_2/n_3$. 

The bottom frame of Fig.~\ref{fig.2b-mmr} also shows evidence of 2P-MMRs between non-adjacent 
planets. The diagonal curve starting from $(n_1/n_2,n_2/n_3) \simeq (2.15,1.35)$ down to 
$(n_1/n_2,n_2/n_3) \simeq (1.35,2.15)$ marks the location of the 3/1 resonance between $m_1$ 
and $m_3$. Although other similar commensurabilities exist in the plot, they are either weaker 
(and therefore difficult to visualize) or are located for period ratios closer to unity and 
drown in the red region of the map. As shown by \cite{Delisle.2017}, mean-motion resonances 
between non-adjacent planets may play an important role in generating new stable fixed points 
for 3-planet resonances.

\begin{figure}
\centering
\includegraphics*[width=0.95\columnwidth]{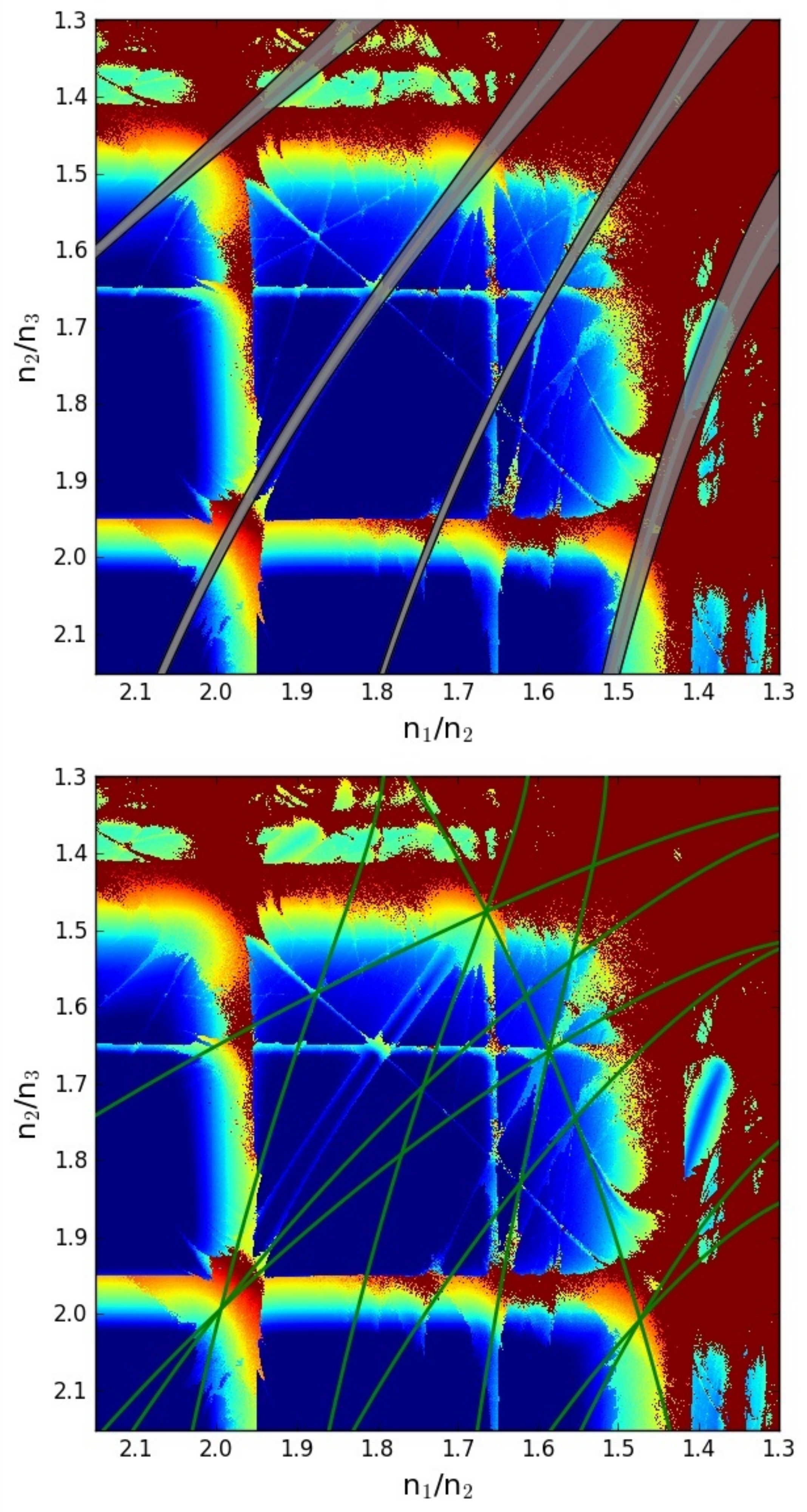}
\caption{{\bf Top:} Location and libration widths of zero-order (i.e. Laplace) 3-planet 
resonances in the dynamical map described in Fig.~\ref{fig.2b-mmr}. From left to right, the 
commensurabilities correspond to $n_1-4n_2+3n_3 = 0$, $n_1-3n_2+2n_3 = 0$, $2n_1-5n_2+3n_3 = 0$ 
and $n_1-2n_2+n_3 = 0$. {\bf Bottom:} Green lines show the position of several first-order 
3-planet resonances.}
\label{fig.3b-mmr}
\end{figure}

\subsection{3-Planet Mean-Motion Resonances (3P-MMR)}

The planets will lie in the vicinity of a 3-planet mean-motion resonance (hereafter 3P-MMR) if 
their mean motions satisfy the linear equation \eqref{eq.14} with $j_i \ne 0 \; \forall i$. We 
can re-write the resonance relation as
\be
p n_1 - (p+q-s) n_2 + q n_3 \simeq 0 \hspace*{0.5cm} ; \hspace*{0.5cm} {\rm with} 
\hspace*{0.2cm} p,s,q \in {\cal Z}.
\label{eq.15}
\ee
The sum of the index is equal to $s$, whose absolute value gives the order of the 3P-MMR. 
Zero-order 3-planet commensurabilities, also referred to as Laplace resonances, correspond to
$s=0$. All exoplanetary systems currently associated to multi-resonant configurations (e.g. 
GJ876, Kepler-60, Kepler-80, Kepler-223) lie in Laplace resonances, as are the well known 
Galilean satellites. So far, only the outer planets of our own solar system appear to be 
affected by high-order 3P-MMRs \citep{Murray.Holman.1999, Guzzo.2005, Guzzo.2006}, possibly 
leading to chaotic motion in Giga-year timescales. 

The location of 3P-MMRs in the dynamical map define curves given by the functions
\be
\biggl( \frac{n_2}{n_3} \biggr)^{-1} = \frac{(p+q-s)}{q} - \frac{p}{q} \biggl( \frac{n_1}{n_2} 
\biggr)
\label{eq15}
\ee
%\citep[e.g.][]{Gallardo.etal.2016}. 
As in the case of 2-planet commensurabilities, these 
relations are given in mean variables and must be transformed to osculating elements before 
plotting them in the representative plane of initial conditions. 

The top frame of Fig.~\ref{fig.3b-mmr} shows the location and libration width (for zero 
eccentricity) of several Laplace resonances ($s=0$). Although an infinite number of Laplace 
resonances exist in the plane, we only plotted those MMRs that led to appreciable values of 
$max(\Delta a)$ during the integration timescale. Although this is not a rigorous criterion, 
these should be the most relevant commensurabilities liable to affect the dynamics of planetary 
systems, at least for the mass values considered here. The libration widths were calculated with 
the analytical model by \cite{Quillen.2011} and show a very good agreement with the structure 
of the dynamical map, although the numerical simulations seem to indicate larger libration 
widths. As shown by \cite{Quillen.2011} (see also \cite{Gallardo.etal.2016}), Laplace 
resonances have a very week dependence with the eccentricities and both branches of the 
separatrix are clearly noticeable for circular orbits. 

The map also shows evidence of first and higher-order 3P-MMRs. The locations of the most 
relevant first-order commensurabilities are plotted as green curves in the bottom frame of 
Fig.~\ref{fig.3b-mmr}. As with their zero-order cousins, most curves have a positive gradient in 
the mean-motion-ratio plane (i.e. $\partial (n_2/n_3)/\partial (n_1/n_2) > 0$); the opposite 
occurs when $q < 0$. The only member of this set plotted here corresponds to the $n_1 + n_2 - 
n_3 = 0$ resonance. 

We can define two different types of 3P-MMRs. If the sub-systems $m_1$-$m_2$ and $m_2$-$m_3$ 
are both in 2-planet resonances such that $pn_1 - k_1n_2 = 0$ and $k_2n_2 - qn_3 = 
0$, then the difference between both will also be zero: $pn_1 - (k_1+k_2)n_2 + qn_3 = 0$. In 
this case, the 3-planet resonance will only be a consequence of the overlap of two independent 
2P-MMRs \citep{Morbidelli.2000} and the dynamics will still be dominated by the individual 
resonant terms stemming from the first-order normal form, and not by the second-order 
perturbation terms modeled by \cite{Quillen.2011}. We refer to such a configuration as a {\it 
double resonance}. The three outer planets of the Gliese 876 system lie in such a double 
resonance, where three of the four two-planet critical angles librate leading to a libration of 
the Laplace angle \citep{Marti.etal.2013}. 

The opposite case occurs when the 3P-MMR condition $p n_1 - (p+q-s) n_2 + q n_3 = 0$ is 
satisfied without the individual planetary pairs exhibiting resonant motion. Following 
\cite{Gozdziewski.etal.2016}, we refer to such a configuration as a {\it pure 3-planet 
resonance}. Only in these cases are resonant models constructed from the Hamiltonian 
second-order normal forms valid, since it is expected that the first-order terms should have 
short periods and near-zero average values. However, as discussed by \cite{Gallardo.etal.2016}, 
this is not always the case and the domain of validity of second-order models 
\citep[e.g.][]{Quillen.2011} may be a strong function of the eccentricities.

\begin{figure}
\centering
\includegraphics*[width=0.9\columnwidth]{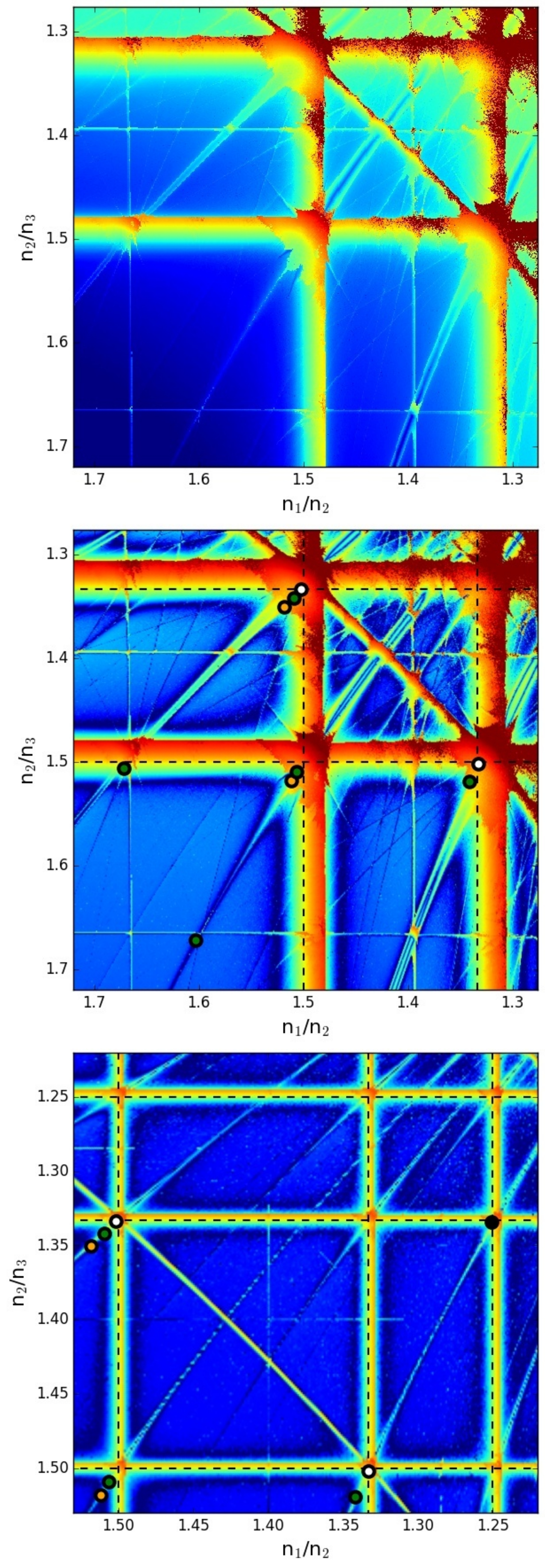}
\caption{{\bf Top:} $max(\Delta a)$ dynamical map for equal-mass planets with 
$m_i=30 m_{\oplus}$, orbiting a central star with $m_0 = 1 m_\odot$. {\bf Middle:} Same as 
above, but after subtracting the amplitudes of the short-period variations. {\bf Bottom:} 
Dynamical map for $m_i=3 m_{\oplus}$, also without short-period variations. Filled circles 
in the two lower plots show the location of four exoplanetary systems: Kepler-223 (white), 
Kepler-60 (black), Kepler-80 (orange) and TRAPPIST-1 (green). Nominal location of first-order 
2P-MMRs are identified with horizontal and vertical dashed lines.}
\label{fig.mapa_fondo}
\end{figure}

Fig.~\ref{fig.mapa_fondo} shows additional dynamical maps, this time constructed for lower 
planetary masses and zooming into mean-motion ratios. For the top and middle graphs we 
adopted masses $m_i = 30 m_{\oplus}$ while in the bottom graph we used $m_i = 3 m_{\oplus}$.
While the strongly chaotic and unstable regions are no longer present in these plots, the 
background value of $max(\Delta a)$ shows a significant increase as the mean-motion ratio 
approaches unity. This pronounced color gradient is caused by the increasing amplitude of 
short-period variations and complicates the identification of the resonant structures in 
different regions of the plane. While we could eliminate this effect applying a low-pass digital 
filter on the output of the numerical integrations, this would have implied an unnecessary 
increase in the computing time. We then opted for a simpler, and more interesting alternative 
method.

The middle frame of Fig.~\ref{fig.mapa_fondo} repeats the top graph, but where we subtracted 
the short-period amplitudes 
\be
\Delta a_i = \frac{4 L_i}{\mu_i {m'_i}^2} \Delta L_i  \hspace*{0.5cm} ; \hspace*{0.5cm} 
(i=1,2,3) , 
\label{eq.16}
\ee
where $\Delta L_i$ are given by expressions \eqref{eq.10}. The result effectively reduces the 
differential background value allowing for a much clearer picture of the structures of the 
representative plane defining the long-term dynamical evolution. The complex web of resonances 
are now enhanced and stand out in all the different regions of the mean-motion ratio plane.

The bottom frame shows a similar map, this time drawn for planetary masses $m_i = 3 
m_\oplus$, and again after removing the short-period variations. Compared to the intermediate 
masses (middle plot), as well as to the map discussed in Fig.~\ref{fig.2b-mmr}, the change from 
mean to osculating elements is much less pronounced, leading to less deformed structures closer 
to the nominal value of the mean-motion ratio. The degree of chaoticity (or semimajor axis 
excitation) is also significantly reduced, although the same is noted for the resonance 
strengths/widths.

\begin{figure}
\centering
\includegraphics*[width=0.95\columnwidth]{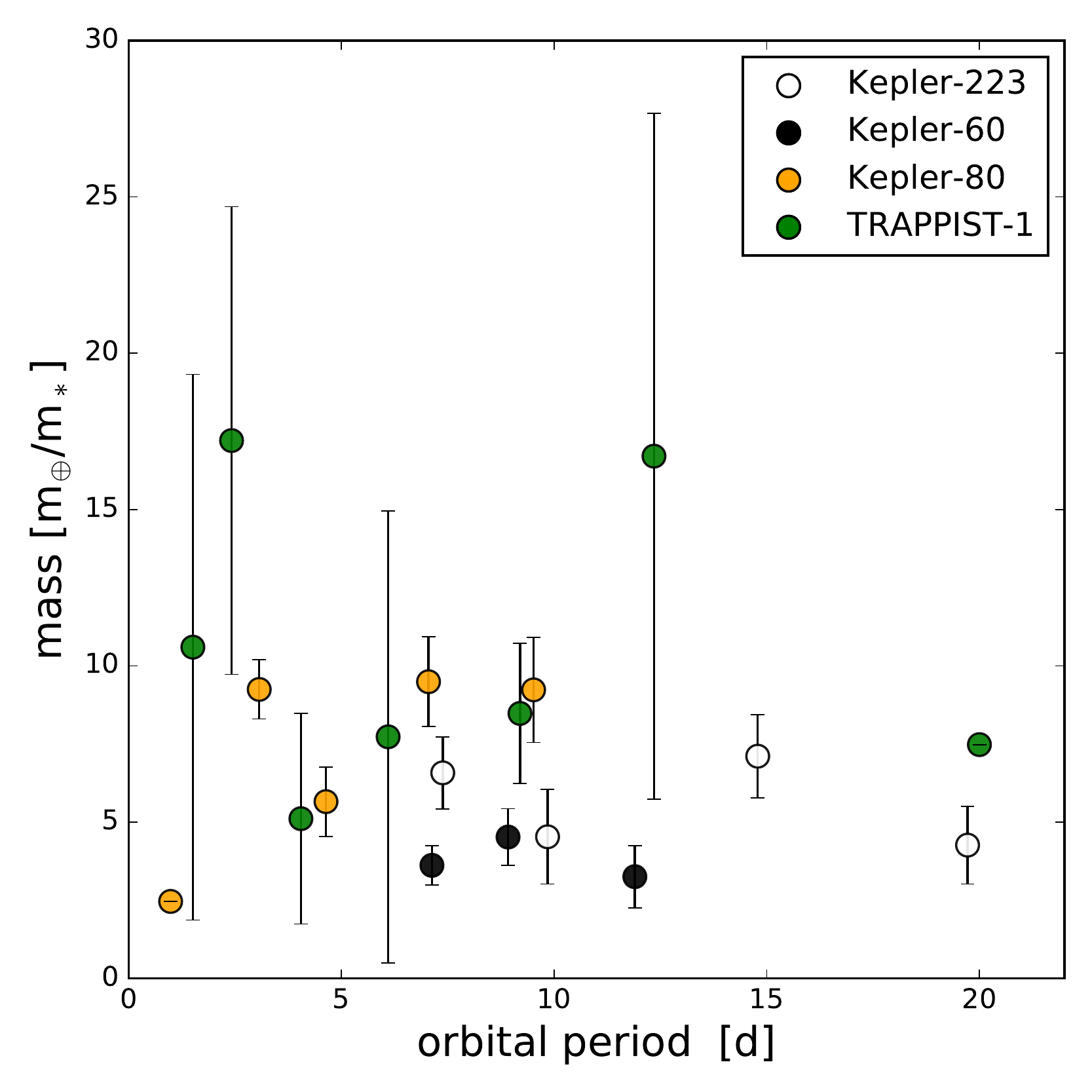}
\caption{Estimated planetary masses as function of orbital period, for several resonant 
multi-planetary systems located close to the host star. Color code is the same as used in the 
previous figure. Masses are given in units of earth-mass divided by stellar mass.}
\label{fig.mp_kepler}
\end{figure}

\subsection{Known Systems in Double Resonances}

The two lower-plots of Fig.~\ref{fig.mapa_fondo} also show the current location of four
close-in multi-planet systems whose dynamics is believed to be dominated by 3-planet resonances. 
The color code employed to identify each system is described in the caption, while the 
estimated masses and orbital period distribution are shown in Fig.~\ref{fig.mp_kepler}. The 
masses for both the inner planet of Kepler-80 and the outer body of TRAPPIST-1 are very 
uncertain and thus these data have been plotted without error bars. Calculated values of
$m_i/m_*$ seem to cover the interval between $\sim 3-30 \; m_\oplus/m_{\boldsymbol *}$, thus the 
general qualitative features of their dynamics should correspond to the middle and lower plots 
of Fig.~\ref{fig.mapa_fondo}. 

While all these systems appear located in double resonances, we can separate them in two 
distinct groups. The first is comprised of Kepler-60 and Kepler-223, whose location in the 
mean-motion ratio plane shows no appreciable offset with respect to the nominal location of the 
double resonances. In the case of Kepler-60, this proximity may be biased since the orbital fit 
process employed by \cite{Gozdziewski.etal.2016} assumed resonant motion as a proxy. 
However, this is not the case of Kepler-223, where the libration of the 3-planet Laplace angles 
has recently been measured from TTV data \cite{Mills.etal.2016} and whose proximity to exact 
resonance appears certain.

TRAPPIST-1 and Kepler-80, representatives of the second group, show a significant displacement 
with respect to the double resonance, although all the 3-planet sub-systems are well aligned 
with location of the zero-order 3P-MMRs. The orbital period distribution of these systems 
(Fig.~\ref{fig.mp_kepler}) shows that both are much closer to their host stars than the members 
of the first group, thus more susceptible to tidal evolution. Depending on the number of 
librating 2-planet resonance angles, \cite{Batygin.Morbidelli.2013} and \cite{Papaloizou.2015} 
proposed that some systems within double resonances could evolve by tidal effects preserving the 
libration of the Laplace angle. Specifically, numerical simulations of Kepler-80 by 
\cite{MacDonald.etal.2016} showed how tidally-induced divergent migration may have lead to final 
orbital architectures similar to the observed system, characterized by large displacements from 
nominal 2P-MMR while preserving libration of the Laplace angles. 
\begin{figure}
\centering
\includegraphics*[width=\columnwidth]{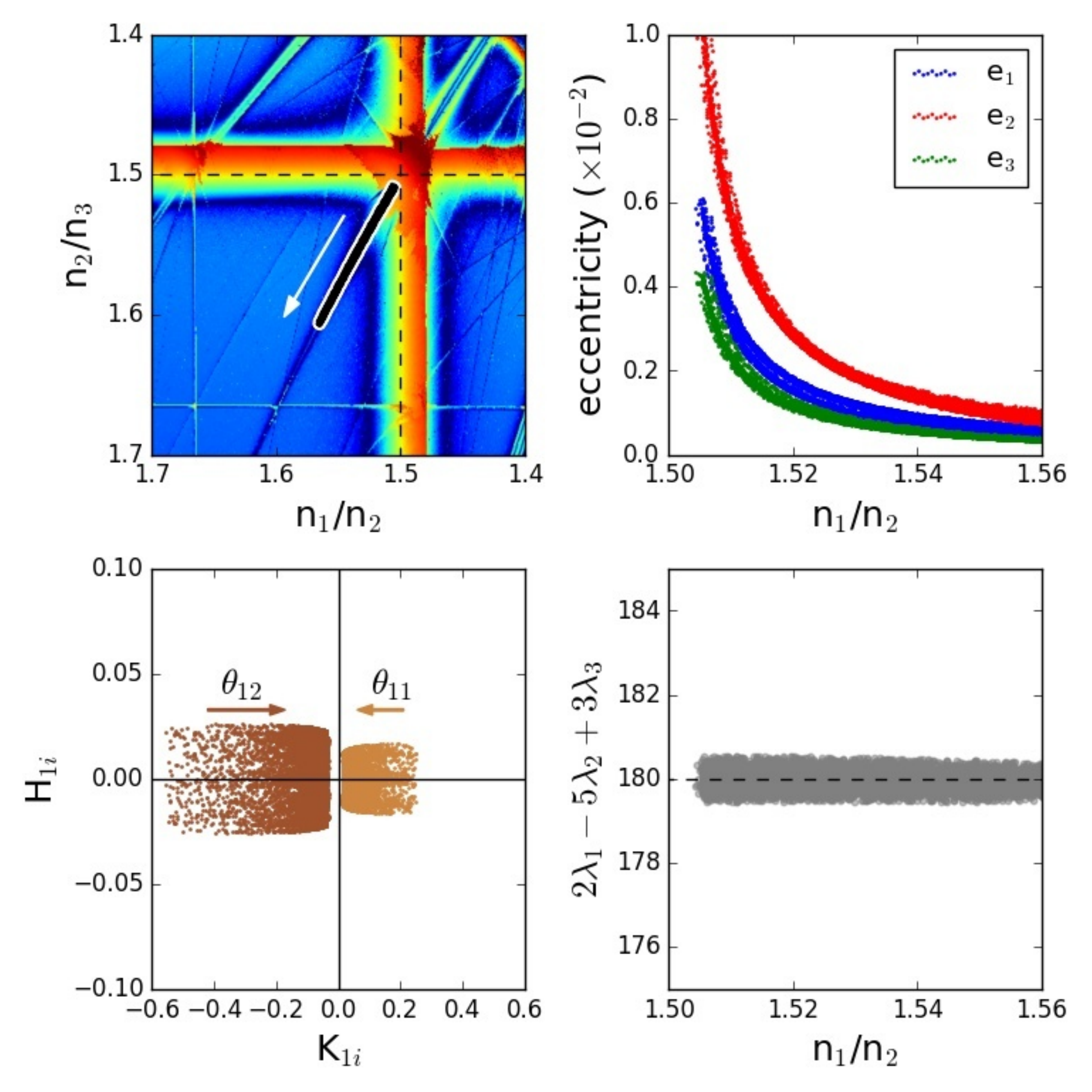}
\caption{Tidal evolution of a fictitious 3-planet system, with masses $m_1 = 7 \; m_\oplus$ , 
$m_2 = 10 \; m_\oplus$ and $m_3 = 15 \; m_\oplus$, and initially trapped in a 3/2 double 
resonance. {\bf Top Left:} Superimposed to the dynamical map, the black-over-white line 
shows the evolutionary track of the system, leading towards larger mean-motion ratios but 
following the 3P-MMR $2n_1-5n_2+3n_3=0$. {\bf Top Right:} Eccentricities as function of 
$n_1/n_2$. {\bf Bottom Left:} Behavior of the regular canonical variables $(K_{1i},H_{1i}) = (2 
S_i)^{1/2} (\cos{\theta_{1i}},\sin{\theta_{1i}})$, for the resonant angles of the inner pair: 
$\theta_{1i} = 3 \lambda_2 - 2 \lambda_1 - \varpi_i$. Arrows indicate direction of the 
evolution. Similar behavior is seen for the pair $(K_{2i},H_{2i})$. {\bf Bottom Right:} 
Resonant angle of the Laplace resonance as function of 
$n_1/n_2$.}
\label{fig.laplace}
\end{figure}

To understand how tides affect the distribution of 3-planet resonance chains in the 
$(n_1/n_2,n_2/n_3)$ plane, Fig.~\ref{fig.laplace} shows the tidal evolution of a fictitious 
system comprised of three planets orbiting a $m_* = 1 \, m_\odot$ central star. Initial 
conditions were taken from the final state of a prior simulation of resonance capture, and 
correspond to a very small amplitude libration of all 2-planet resonant angles:
\be
\begin{split}
\theta_{11} &= 3 \lambda_2 - 2 \lambda_1 - \varpi_1 \hspace*{0.5cm} ; \hspace*{0.5cm}
\theta_{12}  = 3 \lambda_2 - 2 \lambda_1 - \varpi_2 \\
\theta_{22} &= 3 \lambda_3 - 2 \lambda_2 - \varpi_2 \hspace*{0.5cm} ; \hspace*{0.5cm}
\theta_{23}  = 3 \lambda_3 - 2 \lambda_2 - \varpi_3 . \\
\end{split}
\label{eq.17}
\ee
The tidal evolution was simulated using the classical equilibrium tide model 
\citep{Mignard.1979} incorporating the precession and dissipation terms into an N-body code 
\cite[e.g.][]{Beauge.Nesvorny.2012}. Since the graphs present correlations between different 
projections of the phase space, and not variables as a function of time, the results are 
independent of the tidal parameters, as long as the evolutionary timescales are adiabatic with 
respect to the librational periods. 

Starting from $(n_1/n_2,n_2/n_3) \simeq (1.504,1.509)$, the divergent migration increased both 
mean-motion ratios driving the system away from the double resonance (upper left-hand frame). 
However, the rates of change are not independent but constrained by the Laplace resonance. As 
found previously by \cite{Papaloizou.2015} and \cite{MacDonald.etal.2016}, the zero-order 3P-MMR 
stemming from the double resonance acts as a trench through which the system evolves. As seen 
in the lower right-hand frame, the corresponding Laplace resonant angle $\phi = 
2\lambda_1-5\lambda_2+3\lambda_3$ librates around $\phi = 180^\circ$ with a very small 
amplitude with no discernible linear deviation.

\begin{figure*}
\centering
\includegraphics*[width=2\columnwidth]{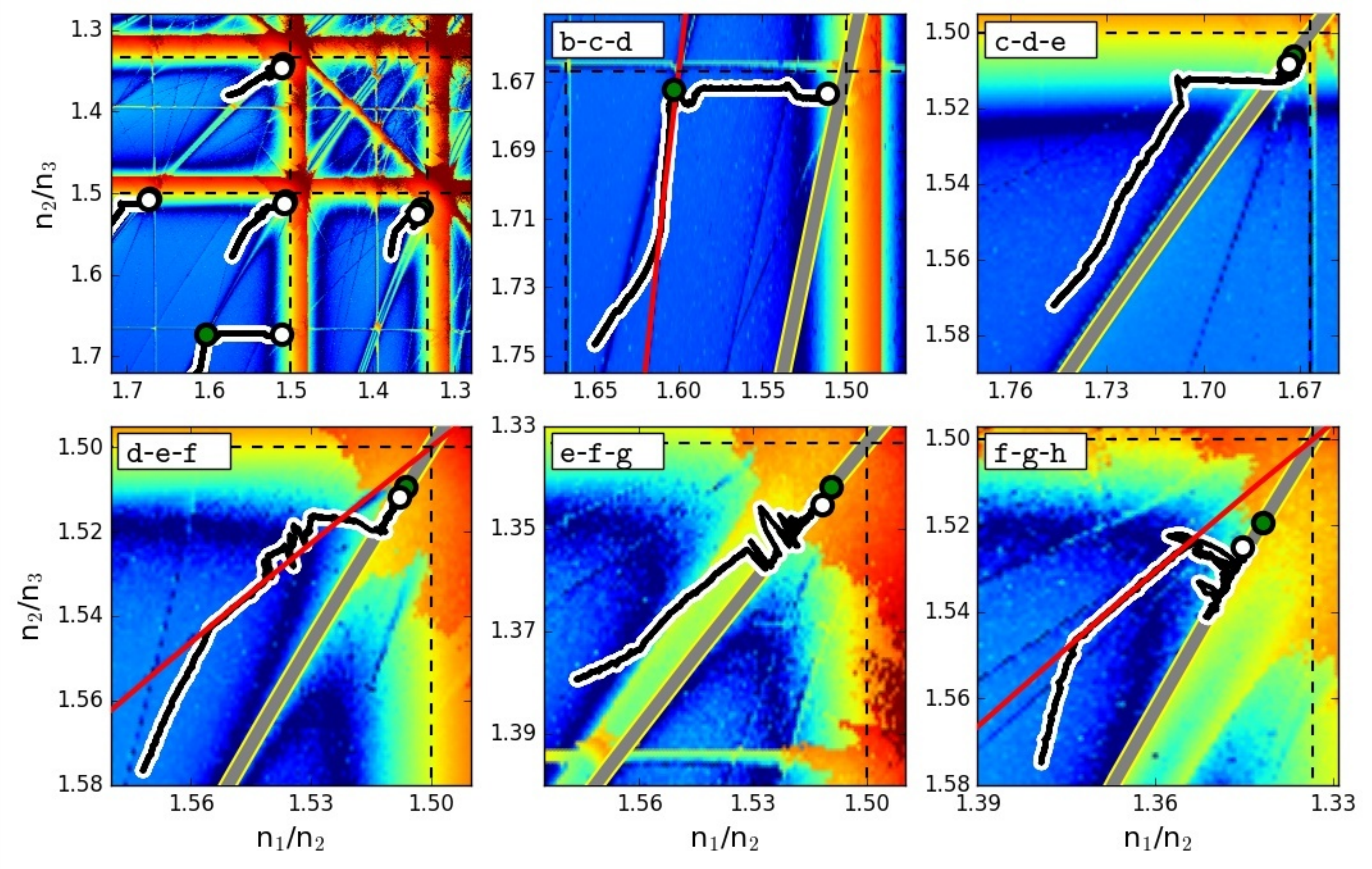}
\caption{N-body simulation of type-I planetary migration in a TRAPPIST-1-like system. Top 
left-hand plot shows the evolution of the mean-motion ratios of consecutive three-planet 
sub-systems, with white filled circles showing the final configuration and the observed 
planets in green. Black-over-white trails are the evolutionary paths followed by each 
sub-system from the initial conditions to the final equilibrium points. All other frames 
show detailed zooms in the vicinity of each sub-system, starting from the triplet 
(\texttt{b}-\texttt{c}-\texttt{d}) (upper middle graph, see inlaid legend) down to 
(\texttt{f}-\texttt{g}-\texttt{h}) in the lower right-hand frame. Relevant 2-planet resonances 
are identified by dashed black lines, Laplace 3P-MMRs are shown with broad gray curves, while 
first-order 3P-MMRs are indicated in red. Numerical data were filtered to eliminate 
short-period and resonant terms in order to reduce spreading, thus the locations of the 
resonances are drawn in mean (not osculating) variables. To guarantee convergent migration, 
planetary masses $m_i/m_*$ were taken equal to $16$, $17$, $17.6$, $18.4$, $19.5$, $21$ and 
$24$, all in units of $m_\oplus/m_\odot$. The numerical integration was stopped when the 
system achieved a steady-state configuration, corresponding to $\sim 5 \times 10^5$ orbital 
periods of the outer planet at its initial semimajor axis.}
\label{fig.trappist}
\end{figure*}

Although we expected the 2-planet resonant angles to circulate once the system increased its 
offset with respect to the double resonance, the bottom left-hand plot shows this is not the 
case. The tidal evolution follows the ACR-loci of solutions \citep{Beauge.etal.2006, 
Michtchenko.etal.2006}, leading to a monotonic decrease in the eccentricities (upper right-hand 
plot) with damped amplitudes of the secular modes. 

While these librations are kinematic and the motion is no longer encompassed by the resonant 
separatrix, the 2-planet resonant terms still seem to be important in defining the dynamics of 
the system, even far from their nominal locations. This raises the issue of the relative weight 
between the pure 3-planet resonant terms \citep{Quillen.2011} and the 2-planet 
perturbations in 
defining the long-term dynamics of the system. Perhaps the underestimation of the analytical 
estimations of the libration width for Laplace resonances (top frame of Fig.~\ref{fig.3b-mmr}) 
is not due to intrinsic limitations in the second-order normal form, but to the first-order 
contributions which were not included.

\section{Resonant Capture in 3P-MMRs}
The multi-resonant extrasolar systems discussed in the previous section are believed to have 
attained their current configuration as a consequence of a smooth planetary migration with the 
primordial gaseous disk. Since their masses $m_i/m_*$ are small, we expect the orbital decay to 
have been dominated by a Type-I migration \citep[e.g.][]{Ward.1997}. 

While in 2-planet systems planet-disk interactions drive the mean-motion ratio to a 
resonance lock in 2P-MMRs, in 3-planet cases the differential migration (i.e. mean motion 
ratio) is only stalled when the complete system is trapped in two independent MMRs. In the 
examples analyzed above, all captures appear to be 2-planet resonances. Thus, all 2-planets 
commensurabilities do not appear to be pure but double resonances.

\subsection{The Case of TRAPPIST-1}

In this scenario, the current orbital configuration of planets \texttt{b}-\texttt{c}-\texttt{d} 
of TRAPPIST-1 (see middle plot of Fig.~\ref{fig.mapa_fondo}) looks curious. According to 
\cite{Gillon.etal.2017}, this sub-system is located in a double resonance identified by 
$(n_1/n_2,n_2/n_3) = (8/5,5/3)$ and thus corresponding to high-order 2-planet resonances. While 
the dynamical map shows evidence of the $5/3$ commensurability, no indication is observed of
the third-order $8/5$ 2P-MMR. However, we do notice a diagonal strip intersecting the observed 
location of \texttt{b}-\texttt{c}-\texttt{d} corresponding to the first-order 3P-MMR $3n_1 - 
6n_2+2n_3=0$. We then ask what role may 3-planet resonances have played in the trapping of 
these planets and whether the evolutionary tracks of the system may have actually followed 
3P-MMRs instead of the traditional 2-planet counterparts.

In an attempt to see some light into this issue we performed a series of N-body 
simulations of type-I planetary migration of TRAPPIST-1-like systems. Instead of introducing an 
ad-hoc exterior force acting only on the outer planet \citep[e.g.][]{Tamayo.etal.2017}, we 
adopted the analytical prescription of \cite{Tanaka.etal.2002} and \cite{Tanaka.Ward.2004}, 
incorporating the partial preservation of the angular momentum suggested by 
\cite{Goldreich.Schlichting.2014}. Full equations of motion and further details of the 
resulting N-body code may be found in \cite{Ramos.etal.2017} and \cite{Zoppetti.etal.2018}. 
Both tidal evolution and relativistic effects were neglected in these simulations.

Since convergent migration required planetary masses increasing with orbital distance, we 
assumed $m_b = 16$, $m_c = 17$, $m_d = 17.6$, $m_e = 18.4$, $m_f = 19.5$, $m_g = 21$ and $m_h = 
24$, all in units of $m_\oplus/m_\odot$. Although these values are arbitrary, they are more or 
less consistent with the estimated masses and uncertainties shown in Fig.~\ref{fig.mp_kepler}. 
We assumed a thin flat laminar disk with $H_0=0.05$ and a surface density profile $\Sigma(r) = 
\Sigma_0 r^{-\sigma}$ with $\sigma=1/2$ and $\Sigma_0 = 50$ gr/cm$^2$. This low surface density 
led to a characteristic migration timescale of $\tau_a \sim 10^5$ years, probably much higher 
than expected for a MMSN but practically equal to that assumed by \cite{Tamayo.etal.2017}. 

Initial conditions were chosen with eccentricities $e_i=0.01$ and all angles equal to zero; 
semimajor axes placed the planets outside (but not very close to) the observed resonance 
locations. By modifying the planetary masses (i.e mass ratios) we were able to generate 
evolutionary tracks in the $(n_1/n_2,n_2/n_3)$ plane with any desired angle, and thereby choose 
which would be the first resonance encountered by each sub-system. This degree of 
freedom contrasts with the approach adopted by \cite{Tamayo.etal.2017} where the sub-systems 
always started migrating following vertical lines in the mean-motion ratio plane. 

Results of a typical run are shown in Fig.~\ref{fig.trappist}, superimposed to the dynamical 
map obtained for $m_i = 30 m_\oplus$. The top left-hand plot shows a global view, with the 
evolutionary trails of the migration in black-over-white lines, while the final configuration 
is highlighted in white filled circles. The current positions of the system is shown in green, 
although in most cases these practically coincide with the simulated system and are virtually 
unseen. The only triplet we were not able to reproduce consists of planets 
(\texttt{b}-\texttt{c}-\texttt{d}) for which our N-body integration ultimately led to a capture 
in $(n_1/n_2,n_2/n_3) = (3/2,5/3)$. 

The remaining plots of Fig.~\ref{fig.trappist} focus on the migration of the different 
sub-system triplets, identified by inlaid legends in the upper left-hand corners. Each will be 
discussed below.

\begin{itemize}

\item
{\bf Planets (\texttt{b}-\texttt{c}-\texttt{d}):} Starting from the lower left-hand end of the 
plot, the 3-planet sub-system approaches and is trapped in the $3n_1 - 6n_2 + 2n_3 = 0$ 
first-order 3P-MMR (red line), thereafter following its trail up to the observed position of the 
real system and the 2-planet resonance $n_2/n_3 = 5/3$. Notice no indication of the $n_1/n_2 = 
8/5$ in the dynamical map. Although the simulated system is temporarily trapped in a location 
close to the observed planets, it is eventually ejected and follows the $n_2/n_3 = 5/3$ 
commensurability until finally resting in $(n_1/n_2,n_2/n_3) = (3/2,5/3)$. The broad gray 
line corresponds to the Laplace resonance $4n_1 - 9n_2 + 5n_3 = 0$. 

Even after several attempts with different masses and disk parameters, we were unable to find 
any cases of a permanent stable capture in the double resonance $(n_1/n_2,n_2/n_3) = 
(8/5,5/3)$. This apparent inconsistency with the results of \cite{Tamayo.etal.2017} could be due 
to differences in the modeling of the planetary migration, or perhaps a more thorough 
exploration of the parameter space is needed.
\vspace*{0.2cm}

\item
{\bf Planets (\texttt{c}-\texttt{d}-\texttt{e}):}
This is a straight-forward case. The direction of relative migration avoids any significant 
pure 3P-MMR and the outer pair is initially in the $n_2/n_3 = 3/2$ resonance. Later, 
migration 
follows this commensurability until reaching the double resonance $(n_1/n_2,n_2/n_3) = 
(5/3,3/2)$ stopping very close to the current location of the observed planet triplet 
configuration. 
\vspace*{0.2cm}

\item
{\bf Planets (\texttt{d}-\texttt{e}-\texttt{f}):}
After an initial migration in a secular configuration, the system is trapped in the first-order
$2n_1 - 7n_2 + 6n_3 = 0$ pure 3P-MMR (red line), following its trail until reaching the 
vicinity of the double resonance where the trajectory begins to exhibit irregular oscillations. 
At one point the system leaves the first-order resonance and is trapped in the strong Laplace 
commensurability defined by $2n_1 - 5n_2 + 3n_3 = 0$ (broad gray line), where it continues to 
migrate until reaching a final destination very close to the actual planets. The capture into 
the pure zero-order 3P-MMR does not seem to follow a smooth transition but seems 
consequence of 
small-scale scattering caused by perturbations onto the first route followed by the system.
\vspace*{0.2cm}

\item
{\bf Planets (\texttt{e}-\texttt{f}-\texttt{g}):}
Contrary to the previous case, this sub-system appears to suffer a smooth capture into the pure 
Laplace resonance $n_1 - 3n_2 + 2n_3 = 0$ (broad gray line) early in its migration, although we 
cannot rule out a possible first-order 3P-MMR guiding the first part of the integration. 
However, we were unable to find a commensurability relation of this kind that was sufficiently 
strong to explain the transition between the initial condition and the Laplace resonance. 
\vspace*{0.2cm}

\item
{\bf Planets (\texttt{f}-\texttt{g}-\texttt{h}):}
The final and most interesting example is the sub-system composed of the three outer-most 
planets. At first hand, the overall evolution follows closely that of 
(\texttt{d}-\texttt{e}-\texttt{f}), with an initial capture in the $3n_1 - 8n_2 + 6n_3 = 0$ 
resonance (red line) and later switching over to the Laplace $n_1 - 2n_2 + n_3 = 0$ 3P-MMR.
However, what makes this case particularly noteworthy is the large final offset with respect to 
the nominal values of the double resonance, even larger than the value measured for the 
real planets. However, both the simulated and observed planets show no appreciable 
displacement from the zero-order pure 3-planet commensurability. 

\end{itemize}

While we were unable to completely describe the migration and formation of the full resonance 
chain of the TRAPPIST-1 system, and the present relative location of the three inner planets was 
not obtained, the results of these simulations have shown unexpected insights into the complex 
dynamics of multi-resonant systems. The first conclusion is that 2-planet resonances are not the 
only commensurabilities capable of trapping multi-planet systems. If the migration timescale is 
sufficiently large, first-order pure 3P-MMRs may also lead to capture and guide the 
system towards additional commensurabilities. Another unexpected result is capture into 
Laplace-type resonance, although here it is not clear whether these can be reached through a 
smooth migration or require passage trough a chaotic layer generated by the interaction with 
other resonances. Whatever the explanation, these examples point to a diversity of dynamics 
much richer than previously imagined.

A second and perhaps more important result is the large resonance offset attained by the bodies 
without the need of assuming later stage tidal evolution. The sub-system comprised by planets 
(\texttt{f}-\texttt{g}-\texttt{h}) is probably the best example where our simulation led to 
values significantly displaced with respect to the nominal mean-motion ratios, even larger than 
the observed quantities. Moreover, since this sub-system is the farthest from the central 
star, it is expected that tidally induced divergent migration would be less important in this 
case than for the other planet triplets. Perhaps the explanation does not lie in tidal effects, 
but solely in the resonant dynamics and coupling of the different links involved in the 
resonant chain. 

It is nevertheless necessary to bear in mind that the magnitude of the resonant offset is a 
strong function of the planetary masses, regardless of whether we assume resonant interactions 
or tidal effects. For this reason we do not expect our offsets to be exactly equal to the 
observed values. However, it is compelling to note that the offsets obtained from our 
simulation increase for sub-systems farther from the central star, as also appears to the be 
case of the observed TRAPPIST-1 planets.

\subsection{Resonance Trapping of Fictitious Systems}

Given the rich diversity in resonant captures noted in the previous example, we wished to study 
if other outcomes were also possible. In particular, we wondered whether sufficiently long 
migration timescales in fictitious 3-planet systems could lead to permanent stable captures in 
resonant configurations that are not associated with double resonances between adjacent 
planets. 

We performed a series of N-body simulations similar to that described in the previous 
sub-section, varying planetary masses, initial semimajor axes and the surface density of the 
disk. The corresponding orbital migration timescales were found to lie in the interval $\tau_a 
\in [10^4,10^7]$ years. While fast migrations always led to capture in strong double 
resonances, slower rates of orbital decay yielded a wider range of possibilities. Finally, to 
allow for a more direct comparison with compact multi-planet systems, we restricted the 
masses to $1-30 m_\oplus$. We also adopted $m_0 = 1 m_\odot$ for simplicity.

\begin{figure}
\centering
\includegraphics*[width=\columnwidth]{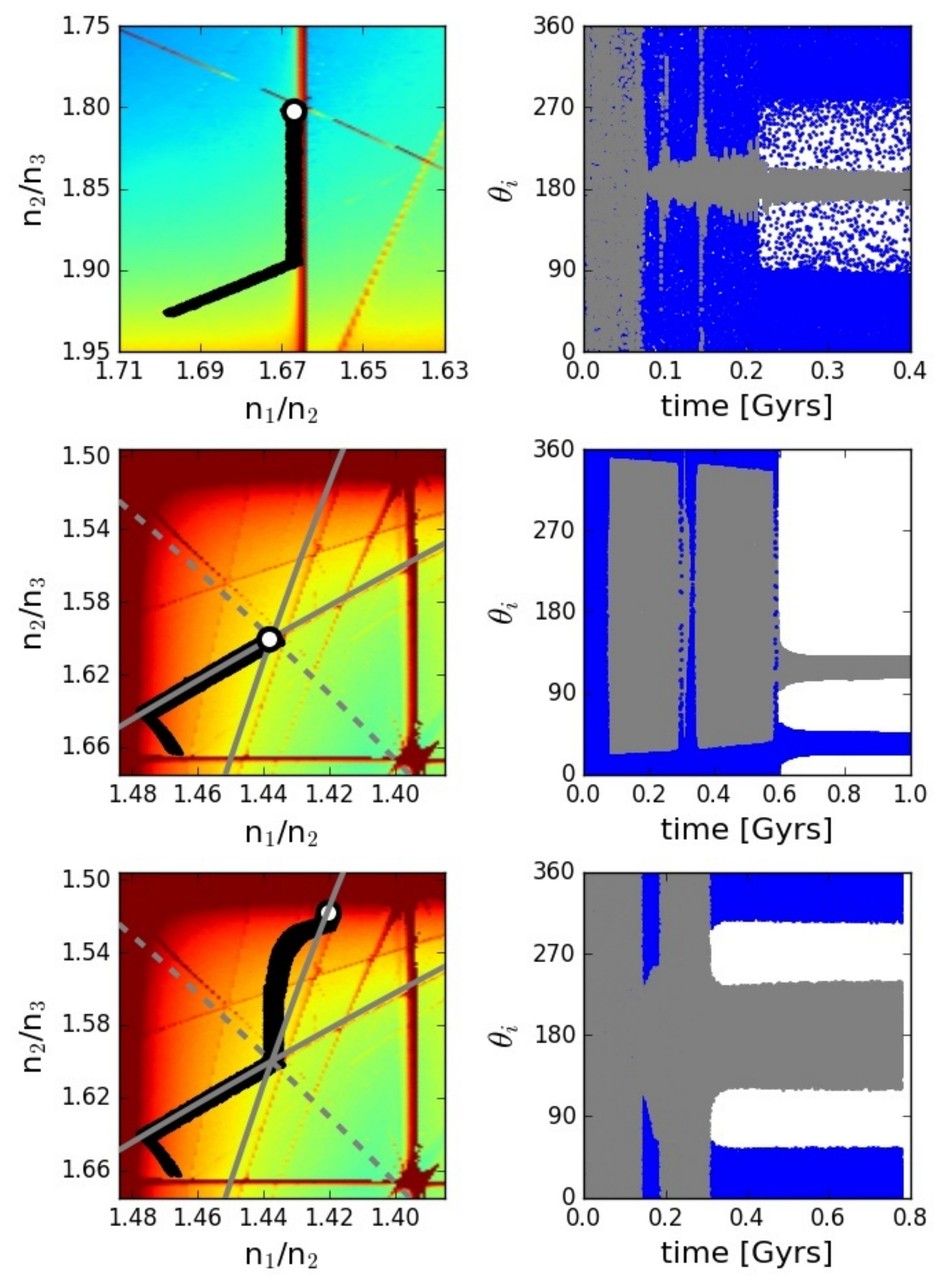}
\caption{Results of three N-body simulations of fictitious 3-planet systems. Black dots 
in the left-hand plots show the evolutionary tracks in the $(n_1/n_2,n_2/n_3)$ superposed over 
the dynamical map constructed for $m_i=30 m_\oplus$. Gray continuous lines show the location of 
the $2n_1-6n_2+5n_3=0$ and $4n_1-7n_2+2n_3=0$ first-order 3P-MMRs, while dashed lines 
correspond to the second-order $2n_1-n_2-3n_3=0$ resonance. The locations of these 
resonances have been plotted in mean elements and thus show a displacement with respect to the 
dynamical map. Right-hand plots show the temporal behavior of resonant angles involved in each 
simulation. See text for details.}
\label{fig.ficticios}
\end{figure}

Fig.~\ref{fig.ficticios} shows the results of three simulations showing diverse outcomes. The 
top two frames correspond to an N-body run with planetary masses $m_1=16$, $m_2=19$ and 
$m_3=22$, all in units of Earth-masses. For the disk surface density profile $\sigma=1/2$, 
these mass ratios guaranteed convergent migration, seen in the dynamical map as an initial 
diagonal evolutionary track leading towards $n_i/n_{i+1} \rightarrow 1$. The surface density of 
the disk at $r=1$ au was chosen equal to $\Sigma_0 = 20$ gr/cm$^2$. 

First the two inner planets are trapped in the 5/3 2P-MMR, after which the system continues to 
evolve vertically until reaching the $n_1 - 3n_3 = 0$ commensurability. This corresponds to a 
3/1 2P-MMR between the inner and outer planet and may be seen in the map as a diagonal line 
crossing the graph in an obtuse angle. Although planetary migration does not stop and all 
semimajor axes continue to decrease, the system arrived at a stable stationary solution with 
eccentricities of the order of $e_i \sim 10^{-3}$ and no further secular change in the 
mean-motion ratios. The right-hand plot shows the temporal behavior of the resonant angles 
$5\lambda_2 - 3\lambda_1 - 2\varpi_2$ (gray) and $3\lambda_3 - \lambda_1 - 2\varpi_1$ (blue). 
Both librate around symmetric values indicating that the system is in fact trapped in an orbital 
configuration in which the inner planet is simultaneously in a 2-planet MMR with the middle and 
outer planet (respectively), but $m_2$ and $m_3$ are not themselves in a resonant motion. 

The two middle frames (left and right) show the result of a second simulation. This time we 
adopted masses $m_1=18.4$, $m_2=19$ and $m_3=22$ (in units of $m_\oplus$), which implies a 
slight increase in the inner mass with respect to the previous case. The aim was to generate an 
initial divergent migration between $m_1$ and $m_2$ and analyze how the full system reacted to 
this non-trivial situation. The surface density of the disk was left unchanged, but the initial 
separations between the planets was reduced in order to study a region of the phase space more 
densely populated by 3-planet resonances. 

As before, the evolutionary track in the plane of mean-motion ratios is depicted in the 
left-hand plot. The initial divergence of the inner planetary pair is stopped as soon as the 
system encounters and is trapped in the $2n_1-6n_2+5n_3=0$ first-order 3P-MMR. From this 
point onwards the corresponding resonant angle $2\lambda_1-6\lambda_2+5\lambda_3 - \varpi_3$ 
begins to librate around an asymmetric center (blue dots in right-hand graph), although it 
suffers a temporary circulation as it suffers a tangential pass through another commensurability 
during its path. The subsequent migration follows the $2n_1-6n_2+5n_3=0$ family until it 
encounters the $4n_1-7n_2+2n_3=0$ resonance. This intersection of two independent 3P-MMRs acts 
as a planetary trap, effectively stalling any additional differential migration. From this 
point onwards the critical angle $4\lambda_1-7\lambda_2+2\lambda_3 + \varpi_3$ also begins to 
exhibit a libration, also around an asymmetric solution, while the eccentricities remain 
only marginally excited at $e_i \sim 10^{-3}$. 

The permanent and dynamically stable capture into two independent first-order 3P-MMR is a 
previously unknown outcome of slow migrations in 3-planet systems. The second-order 3-planet 
resonance marked with dashed lines in the dynamical map did not show any appreciable dynamical 
effects in the system. However, it is interesting to note that 
\be
(4n_1-7n_2+2n_3) - 2(2n_1-6n_2+5n_3) = 5n_2-8n_3
\label{eq.18}
\ee
which implies that a simultaneous libration in both first-order 3P-MMRs will also lead to a 
libration of the two outer planets in the $n_2/n_3 = 8/5$ resonance. This is the same 
commensurability which is believed to dominate planets (\texttt{b}-\texttt{c}-\texttt{d}) of 
TRAPPIST-1. 

The question now is to elucidate which resonances are the cause and which is the consequence. 
At first hand, we would expect that even a third order 2-planet resonance such as the 8/5 
commensurability would be more significant than a first-order 3P-MMR. However, the absence of 
any indication of the 8/5 resonance in the dynamical map raises some doubts. 

Although the intersection of two independent first-order 3P-MMR is always associated to 
2P-MMRs between adjacent planets, many times these are of high order and thus dynamically 
negligible. For example, while the interaction of the 3-planet commensurabilities discussed in 
the middle plot of Figure \ref{fig.ficticios} lead to a 8/5 resonance between $m_2$ and $m_3$, 
the corresponding mean-motion ratio of the inner planetary pair is $n_1/n_2 = 23/16$, a very 
high-order commensurability of dubious influence. Consequently, it is possible that the capture 
process of both TRAPPIST-1 and the fictitious system in Fig.~\ref{fig.ficticios} may actually 
be dominated by first-order 3P-MMRs and not by high-order 2P-MMRs.

The two bottom frames of Fig.~\ref{fig.ficticios} correspond to a third simulation, with 
exactly the same masses and initial conditions as before, but with a higher disk surface 
density: $\Sigma_0 = 40$ gr/cm$^2$. Although the first stages of the migration process are 
similar, the faster migration can no longer be  counterbalanced by the intersection of both 
first-order 3P-MMRs. After a temporary capture, the system passes through and continues to 
evolve towards a more compact configuration. However, after a certain time the planets again 
encounter the $4n_1-7n_2+2n_3=0$ resonance and the capture is repeated, as seen by the behavior 
of the critical angle $4\lambda_1-7\lambda_2+2\lambda_3 + \varpi_3$ (gray dots in the 
right-hand plot). 

The final lap of the evolution follows this resonant family until it encounters the 
2-planet $n_2/n_3 = 3/2$ and all further differential migration stops. The blue dots 
in the right-hand plot show the behavior of $3\lambda_3 - 2\lambda_2 - \varpi_2$, indicating a 
moderate-amplitude libration around zero and a stable orbital configuration. This simulation 
therefore shows a different possible outcome of the migration of 3-planet systems, in which the 
pair of resonances acting as a planetary trap is composed of a first-order 3P-MMR plus a (more 
classical) first-order 2-planet commensurability. 

The three N-body simulations shown in Fig.~\ref{fig.ficticios} show completely different 
outcomes. While in all cases the relative migration is only halted at the intersection of two 
independent resonances, these are not restricted to 2P-MMRs but may include a wide range of 
possibilities. Interestingly, none of these final configurations would be identified as 3-planet 
resonances just from the individual (two planet) mean-motion ratios, but only after a detailed 
analysis of the complete three planet system. The dynamical maps and the identification of 
relevant multi-planet resonances prove important tools to aid in such a search.

\begin{figure}
\centering
\includegraphics*[width=0.99\columnwidth]{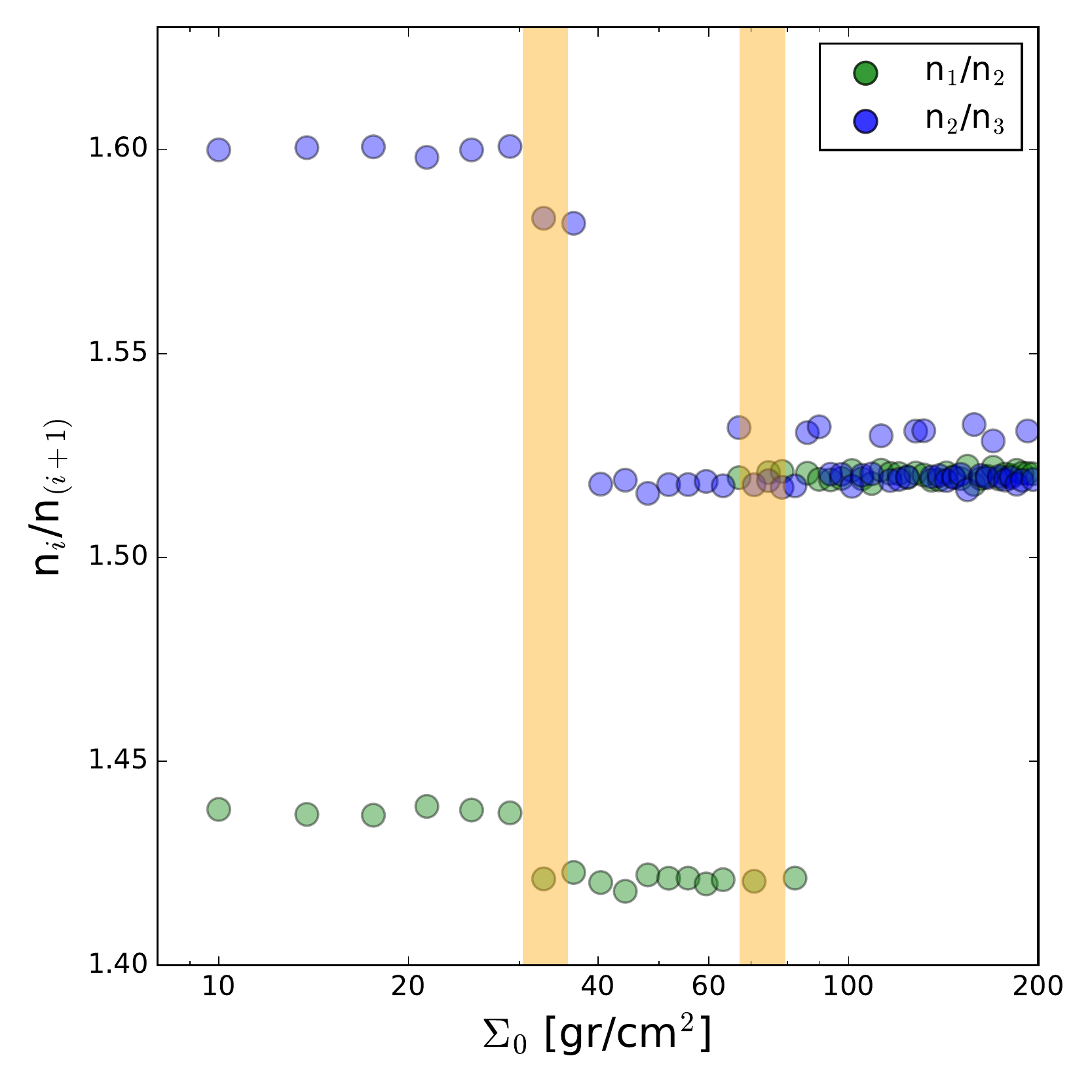}
\caption{Final values of the mean-motion ratios of a series de N-body simulations of migration 
of 3-planet systems with different disk surface densities $\Sigma_0 = \Sigma(r=1)$. Planetary 
masses and initial conditions were taken equal to those discussed in the middle and lower 
frames of Fig.~\ref{fig.ficticios}. Broad orange vertical lines indicate rough limits for three 
distinct outcomes. For $\Sigma_0 \lesssim 30$ gr/cm$^2$, all captures occur in the intersection 
of two first-order 3P-MMRs, while for $\Sigma_0 \gtrsim 80$ gr/cm$^2$ the planets are halted in 
a 3/2 double resonance. For intermediate surface densities the systems evolve towards an 
intersection between a 2-planet and a 3-planet commensurability.}
\label{fig.sigma0}
\end{figure}

Finally, in order to analyze how the final orbital configuration depends on the migration 
timescale, we repeated the previous simulation for a total of 50 values of the disk surface 
density in the interval $\Sigma_0 \in [10,200]$ gr/cm$^2$. For each run we calculated the final 
equilibrium values of $n_1/n_2$ and $n_2/n_3$, plotting their values as function of $\Sigma_0$. 
Results are shown in Fig.~\ref{fig.sigma0}. 

For surface densities $\Sigma_0 \lesssim 30$ gr/cm$^2$, corresponding to migration 
characteristic timescales $\tau_a \gtrsim 1.5 \times 10^5$ years, the system is captured in an 
orbital configuration analogous to that described in the middle plots of 
Fig.~\ref{fig.ficticios}. In other words, the relative orbital decay is stalled by the apparent 
intersection of two independent first-order 3P-MMRs. Since a linear combination of both 
resonant relations yields $5n_2 - 8n_3 = 0$, the two outer planets are also seen to be affected 
by this high-order 2-planet commensurability.

For slightly larger surface densities, leading to $\tau_a$ roughly between $8 \times 10^4$  
and $1.5 \times 10^5$ years, the combined effects of both the $2n_1-6n_2+5n_3=0$ and
$4n_1-7n_2+2n_3=0$ resonances are not strong enough to act as a planetary trap and the system 
evolves towards a new stationary solution involving the $4n_1-7n_2+2n_3=0$ three-planet 
resonance and the $2n_2-3n_3=0$ two-planet commensurability. However, this orbital 
configuration only appears possible for a limited range of disk densities and constitute a 
transition between the low and high density scenarios.

Last of all, for planetary migrations corresponding to $\tau_a \lesssim 8 \times 10^4$ years, 
no 3P-MMR appears sufficiently strong to counteract the dissipative effects and the planets are 
finally captured in a $3/2$ double resonance. All these outcomes were found to be dynamically 
stable and at least one of the resonant angles was observed to librate around a stationary 
point with low-to-moderate amplitudes. 

Of course, the limit between these different regimes depends on the masses of the planets, as 
well as other disk parameters such as the flare index and surface density slope. More complex
physics (e.g. radiative disks or localized dead zones) may also affect these numerical values 
and alter the effective reach of the 3-planet resonances.

\section{Conclusions}

Recent discoveries of compact multi-planet systems have revealed several cases of resonant 
chains (e.g. Kepler-60, Kepler-80, Kepler-223 and TRAPPIST-1) comprised of interlocked 
2-planet and 3-planet commensurabilities. Although these systems are believed to display 
complex dynamical behavior, including the possibility of numerous independent asymmetric 
librational solutions \citep{Delisle.2017}, all 3-planet commensurabilities have so far been 
associated to double resonances and not to pure 3P-MMRs. 

In this paper we have unveiled a more general view of the gravitational interaction of 3-planet 
systems, including a global catalog of mean-motion resonances and possible evolutionary routes 
from secular to resonant configurations. Our study is based on the construction and detailed 
analysis of dynamical maps in the mean-motion ratio representative plane of initial conditions. 
These maps uncovered an extremely rich diversity of possible resonant configurations, including 
zero-order (Laplace-type) and first-order pure 3P-MMRs. Although resonances are dense in 
the representative plane, not all are equally important. In the absence of adequate analytical 
models, these maps allowed us to evaluate their relative strengths and identify which could be 
relevant to the orbital evolution of 3-planet systems. 

While commensurability relations are defined in mean variables, the representative planes of 
initial conditions were chosen in osculating elements. While the difference between both sets is 
usually neglected, in our case it proved important generating a significant shift in the 
position of the resonances with respect to the nominal values. To solve this problem we 
constructed and applied a simple analytical model for the transformation between mean and 
osculating semimajor axes. This model proved vital to properly identify which 3P-MMRs were 
associated to each dynamical feature of the map. As an added bonus, this analytical model 
allowed us to eliminate the background orbital excitations generated by short-period 
perturbations, thus enhancing the long-term dynamical effects throughout the different regions 
of the representative plane.

It is important to stress that the maps were drawn for equal-mass bodies for only three 
specific values of the planetary masses, and thus they are not expected to be exactly the same 
for any other set $(m_1,m_2,m_3)$. Nevertheless, their general features and resonance locations 
should still be qualitatively correct, at least for masses in the same overall range. Thus, even 
if only illustrative, we have extensively used these generic maps as benchmarks in which to 
analyze the dynamical interactions of real and fictitious planetary systems.

The effective strength of first-order pure 3P-MMRs was tested with a series of N-body 
simulations of type-I migration. For fictitious 3-planet systems we found that a complete 
resonance chain may be formed even if the differential migration between some pairs was 
initially divergent. Relative migration was only stalled once the system was trapped in two 
independent mean-motion resonances. For short migration timescales, the intervening 
commensurabilities are 2P-MMRs, such as those associated to Kepler-60, Kepler-80 and Kepler-223. 
However, we also found that slower migration rates lead to a wider range of possibilities, and 
multi-planet systems may be trapped in a combination of 2-planet and pure first-order 
3-planet resonances. Depending on the masses, there always seems to exist an upper limit for the 
disk surface density, under which two independent pure first-order 3P-MMRs may effectively 
trap the system into a permanent and stable configuration not associated to any 2P-MMR.

The possibility of resonant chains not involving 2-planet resonances is intriguing, since such 
a system would not be easily identified as multi-resonant just by plotting the mean-motion 
ratio of adjacent planets. Multi-planet captures such as that depicted in the middle frames of 
Fig.~\ref{fig.ficticios} would also not be associated to a 3-planet resonance. This raises the 
question if the distribution of known multi-planet systems may indeed harbor examples of such 
configurations. We are currently analyzing this possibility, and although no global correlation 
has been found, some individual systems seem promising.

We applied our dynamical maps and migration simulations to the case of TRAPPIST-1 system. 
Starting from initial separations close to but wider tan the current system, we found that most 
planet-triplets halt their relative migration in the double resonance observed today. However, 
the evolutionary routed towards these nesting places were usually guided by first-order 3P-MMRs 
and that some captures into Laplace-type 3P-MMRs were also possible before the 2-planet 
commensurabilities are attained. 

A curious case is that of the inner planets (\texttt{b}-\texttt{c}-\texttt{d}) of TRAPPIST-1, 
believed to lie in the $n_1/n_2=8/5$ and $n_2/n_3=5/3$ double resonance. We could not find 
initial conditions or disk parameters leading to a stable capture into this orbital 
configuration, although some temporary librations were detected for some parameters. However, 
this difference in results with respect to \cite{Tamayo.etal.2017} could be due to 
differences in the migration prescription or the adopted disk parameters. Although our 
dynamical maps were able to detect signatures from a wide range of different resonances, we 
found no evidence of the $8/5$ two-planet commensurability in either $max(\Delta {a})$ nor
$max(\Delta {e})$. Since this is a third-order resonance, its absence could be due to initial 
circular orbits. However, it could also point to a case similar to that shown in the bottom 
plots of Fig.~\ref{fig.ficticios} in which the simultaneous libration of a 2P-MMR and a 
first-order pure 3P-MMR combine to show a libration in the $8/5$ two-planet resonance even 
if this commensurability was not an active ingredient in the capture process. As we showed in 
Fig.~\ref{fig.ficticios}, such a final configuration is possible only for a limited range of 
migration timescales, which could also explain why we were not able to reproduce it in our 
applications to TRAPPIST-1. 

Finally, notwithstanding planets (\texttt{b}-\texttt{c}-\texttt{d}), our tidal-free capture 
simulations of TRAPPIST-1 led to 2-planet resonance offsets similar to those currently observed 
for the real planets. It then appears possible that tidal evolution in multi-resonance systems 
may not have played such an important role as previously believed. However, similar studies in 
other systems (e.g. Kepler-80) are necessary before proposing that these findings are general 
and not restricted to this particular case.

\section*{Acknowledgements}
The authors would like to express their gratitude to the computing facilities of IATE and UNC, 
without which these numerical experiments would not have been possible. We also thank an 
anonymous referee for valuable comments and suggestions that improved the manuscript. This work 
was funded by research grants from SECYT/UNC, CONICET and ANPCyT.

\end{document}